Evaluation d'un simulateur de vissage ilio-sacré percutané

Assessment of a percutaneous iliosacral screw implantation simulator

J. Tonetti[1], L. Vadcard[1], P. Girard[1], M. Dubois[1], P. Merloz[1], J. Troccaz[1]

Corresponding Author's Institution: Orthopaedic and trauma department. Michallon Hospital, BP217, 38043 Grenoble cedex 09

Corresponding Author: Pr Jérôme TONETTI, M.D., Ph.D. Orthopaedic and trauma department. Michallon Hospital, BP217, 38043 Grenoble cedex 09
Mail : JTonetti@chu-grenoble.fr

Short title : Assessment of an iliosacral screwing simulator

Level of evidence : Level III prospective diagnostic study

ABSTRACT

Fundament :

Orthopaedic and trauma surgery training simulation is quite uncommon, however it is a valuable tool to train orthopaedic surgeons and help them plan complex procedures.

Purpose

The aim of that work was to assess the educational efficiency of a percutaneous path simulator under fluoroscopic control applied to the implantation of iliosacral screws.

Material and methods

We evaluated 23 surgeons who inserted a guide-wire in a human cadaver according to a pre-determined procedure. Medical students were defined either as novice or skilled, with or without theoretical knowledge and with or without procedural knowledge. Screw insertion was performed in a human cadaver either without prior training (G1), or after simulator training (G2). Analysis criteria for each surgeon included the number of intraoperative X-rays required during the procedure and a iatrogenic index based on the surgeon's ability to detect any hazardous trajectory.

Results

An average of 13 X-rays was needed by G1 for wire implantation. G2 group required 10 X-rays on average, with use of the simulator. A significant difference was observed in the novice sub-group (N), with 12.75 X-rays on average for G1 and 8.5 for G2. In the sub-group of operators with no procedural knowledge (P-), a significant difference was found since 12 X-rays were required in G1 and 6 in G2. Finally, in the sub-group of operators with theoretical knowledge (T+), a significant difference was found since 16 X-rays were required in G1 and 10.8 in G2. The iatrogenic index was not significantly different.

Discussion

Despite some methodological differences, we were able to demonstrate the simulator's efficiency in familiarizing the operator with the use of a 2-D imaging system to facilitate the procedure in the 3D patient environment. Novice surgeons having a good theoretical anatomical knowledge of the lumbo-sacral joint but no specific knowledge of the surgical technique are those who will best benefit from this tool. The analysis of the training data collected with the simulator will help orientate the surgeon towards non-acquired learning points. The program can easily be extended to any other percutaneous gesture performed under fluoroscopic control.

**Introduction**

Iliosacral screw placement is a useful technique in the fixation of posterior pelvic ring injuries, either sacral bone lesions or sacroiliac joint disruptions [1-4]. The use of fluoroscopy or computer based imaging allows percutaneous iliosacral screw fixation to be performed in the supine position [3, 5, 6] in case of a deteriorated pelvic ring. It ensures an early fixation in polytraumatized patients and significantly reduces the risk of haemorrhagic or infectious complications of the operative site. However, screw placement is not devoid of risks due to the vital structures surrounding the first sacral elements. The lumbosacral nerve trunk in the upper and front part along with the first sacral nerve in the lower and back portion of the sacral wing are exposed to an extra-osseous trajectory of the screws [7]. The inner space of the wing is corridor-shaped with an ovoid transverse section in its narrowest portion of 22 mm on average in its long axis (range 17 to 29 mm) and 11 mm in its short axis (range 9 to 16 mm) [8]. The small dimensions of this narrow section require a very precise trajectory of the screw(s). The reading of the radiographs uses bony landmarks on inlet, outlet and AP views [9, 10] in order to control the trajectory in this narrow bone space. The learning curve of this image-based technique is long and requires much practice through various clinical situations to improve the screw insertion technique [11]. Another adverse effect reported during the first attempts is the patient and surgical team exposure to radiation, which is up to 3.1 minutes in our experience.

The use of computerized surgical simulators is emerging. Computerized surgical simulators were first used for laparoscopy due to the correlation between the operative time and the surgeon's training for some procedures [12]. In the orthopaedic field, knee arthroscopy was the first area to benefit from surgical simulation [13]. Facing the development of percutaneous procedures, a thorough procedural training appears necessary to reduce the operative time and the exposure to ionizing radiation. 3-D computer graphics and virtual reality techniques provide a better understanding of complex 3-dimensional bony structures for proper handling of instruments. Simulators used for the osteosynthesis of proximal femoral fractures performed under fluoroscopic control have shown to be a reliable tool which improves the accuracy of the procedure and reduces the operative time and the need for intraoperative X-rays. [14].

During the European project VŒU IST 1999-13079 « Virtual Orthopaedic European University », we developed a simulator applied to the iliosacral screw placement [15]. This application was chosen by other teams [16]. The aim of our work was to validate this device through real-life application scenarios in a large surgical population.

**Material and methods**

**Evaluated population**

23 trainee orthopaedic surgeons were evaluated while attending an institutional course on pelvic surgery (Cours Bassin AO France – Club Bassin Cotyle, Nice 14-16 Mai 2008, Laboratoire d'anatomie, Pr De Peretti). They were advanced interns or young assistant-senior registrars who did not perform percutaneous iliosacral screw placement on a regular basis. Surgeons came from 21 different hospital centres from 20 different cities. 22 males and 1 female were included in this study.

A questionnaire form was given at the registration for the course, to collect factual information about the studied population. It included non-interpretive closed questions. Each surgeon was identified according to his experience of percutaneous iliosacral screw placement and thus classified as a skilled (D) or novice operator (N) according to the number of screw placements he had observed or performed under the control of a senior : 0 screw placement, 1 screw placement, from 1 to 5 screw placements, more than 5 screw placements. Table 1 displays the distribution mode of N or D features. In this questionnaire form, surgeons were also classified according to their level of theoretical knowledge (T) [surgical indication of screw placement, standard fluoroscopic views, level of anatomical knowledge] and their level of procedural knowledge (P) [recognition of the wire trajectory orientation on an intraoperative radiograph and in a 3D environment, change in the trajectory on a diagram of the pelvis]. An abstract of the questionnaire form is shown in figure 1. The T+ characteristic was attributed to the subjects having obtained a > 80% score to the theoretical questions. The P+ characteristic was attributed to those with a > 80% score to the procedural questions. Three binary characteristics were attributed to each surgeon : N or D ; T- or T+ ; P- or P+.

**Test procedure**

During the course, a technical account was presented to all surgeons to enunciate the principles of percutaneous placement of iliosacral screws and the stages of the procedure. A percutaneous iliosacral screw placement was performed in human cadaver by each surgeon within a group of 4 surgeons, to take advantage of the institutional course. The stages of the screwing procedure were briefly reminded to each group then surgeons successively implanted a 2.5 mm diameter and 300 mm long wire. Implantation was performed using a mallet then a battery operated motor. The procedure was carried out under standard fluoroscopic control (Siremobil Compact L™, Siemens medical solution, Saint-Denis France) fitted with a 30 cm diameter receiver. Handling was performed by a specialized manipulator according to the surgeon's instructions. The number of fluoroscopic views and the intra- or extra-osseous aspect of trial and final trajectories were noted by an independent observer. An iatrogenic index was defined for each operator according to either the intra-osseous or extra-osseous aspect of the trajectories and above all their ability to recognize the hazardous aspect of the extra-osseous screw placement. Five iatrogenic levels were defined and are displayed in table 2. When the surgeon performed intra-osseous trials and final intra-osseous trajectory, the iatrogenic index was of level 1 minimum. When trials were extra-osseous but final trajectory was intra-osseous, the index was 2. When trials were intra-osseous and the final chosen trajectory was extra-osseous, the index was 3. When trials and final trajectory were extra-osseous, the index was 4. Finally, if no trial was performed and when final trajectory was extra-osseous, the index was 5/5 maximum. For each group, the index was calculated by adding the values obtained by each operator of the group.

**Simulator features**

A simulator was supplied to the surgeons to perform virtual iliosacral screwings [15]. This exercise is available to the reader via the following link : http://www-sante.ujf-grenoble.fr/SANTE/voeu/visfran/index.htm The software was connected to a standard PC. This simulator was used to insert a wire on the right side in a 3-D CT image of the pelvis using a VRML language (cortona VRML client™, parallelgraphics Inc., Dublin, Ireland). The 3-D pelvis was hidden by skin (Fig. 2) and placed in the supine position with a skin landmark (cross on Fig.2) corresponding to the lateral view of the sacrum skin. The commands were : <PLACE> in translation or <ORIENTATE> the wire when outside the body, then <PUSH IN> the wire. The operator could make inlet, outlet and AP X-rays by clicking on the corresponding keys (<INLET> ; <OUTLET> ; <AP>) to check the progression of the wire. In accordance with most situations of the operating theatre, two radiographic views were simultaneously visible (the current and the previous one). Previous images could be seen again by clicking on <PREVIOUS> and <FOLLOWING>. The <RETURN> button would then replace the wire outside the pelvis while holding the entry point and orientation in order to be used as a transitory landmark during the desired corrections. Once the operator was satisfied, he could validate the trajectory by pressing the <CONFIRM> button. The simulator then assessed the final trajectory via a comment : Either the trajectory was successful, that is of intra-osseous aspect and sufficient depth, or unsatisfactory in case of antero cranial or postero-caudal penetration or inadequate/excessive wire progression. The screen displayed these comments which were of immediate educational interest to the operator. The number of radiographic controls required during the session was also displayed. The number of trial procedures was indicated. In case of wrong trajectory, the <LESSON> button would redirect the user to the corresponding lesson about iliosacral screw placement using the simulator. Finally, the <HELP> button indicated each button function when passing over and the <INSTRUCTIONS FOR USE> button opened an explanatory note on how to start the exercise. There was no trace of the exercise in the simulator.

Two groups of surgeons were constituted by equally distributing the three characteristics. G1 performed an iliosacral screw placement on human cadaver, without prior training with the simulator. G2 performed a percutaneous iliosacral screw placement on human cadaver, after a 20 mn simulator training.

**Statistical methods**

The chi-square test was used for statistical analysis to compare the number of radiographs and the iatrogenic index in G1 and G2 groups. The analysis took into account the three characteristics identified for each operator. The mean value of surgeons' iatrogenic index in G1 and G2 was compared.

**Results**

The distribution of characteristics in G1 and G2 groups is reported in table 3.

The mean duration of screw placement for each subject was 6.2 minutes, standard deviation (SD) was 2.2 minutes (G1 : mean duration 6.2 minutes, SD 2 minutes; G2 : mean duration 6,2 minutes ; SD 2.7 minutes). The duration of screw placement was not homogeneously distributed when taking into account the following characteristics : novice/skilled, P+/P- and T+/T- However, no significant difference was found within one characteristic.

The overall number of radiographic controls was 159 in G1 that is an average of 13 controls per operator and 117 in G2 that is an average of 10 controls per operator (Table 4). In G1, novice surgeons required more X-rays (mean 12.75) than in G2 (mean 8.5), which is significant. In G1 group, P- operators significantly required more radiographs (mean 12) than in G2 group (mean 6). T+ operators from G2 group significantly required less radiographs (mean 10.8) than those from G1 group (mean 16).

The overall G1 iatrogenic index was similar to that of G2 which is 21 (Table 5). The mean iatrogenic index per operator was 1.75 (range 1 to 5) in G1 and 1.9 (range 1 to 4) in G2. The mean values of N/D, T-/T+ and P-/P+ sub-groups of G1 and G2 could not be statistically compared due to the small sample and the great number of equal ranks. However, skilled operators from G2 demonstrated a lower iatrogenic index than those from G1. Transitory extra-osseous trajectories were observed 5 times in G1 and 8 times in G2.

## Discussion

Our study reveals numerous methodological aspects. The first aspect refers to the way of identifying the evaluated population. The "novice/skilled" characteristic is suggestive, based on a statement which can not be checked. We tried to examine the operator level of knowledge in more detail via a questionnaire form of theoretical and procedural content. This questionnaire was piloted by a population of interns. The form was filled in freely prior to the registration with the possibility to refer to documents. We suspect the low distinctiveness of this questionnaire. Therefore we put the limit of selection of T+ and P+ characteristics to 80% of positive responses in order to reduce the number of false positive and only retain the subjects with definite knowledge. G1 and G2 groups were constituted so that the characteristics could be harmoniously distributed. However, this balance between characteristics is dependent from the initial data capture. We lack controlled criteria that would help know the level of surgical knowledge. This type of simulator could be a valuable tool to evaluate the surgeon through a pedagogical target such as a surgical procedure [14].

The second aspect concerns the carrying out of the screw placement by 4 operators in human cadaver. Surgeons were not evaluated in an isolated manner. They individually inserted the wire but learnt from each other by observing one another within the group and through the final comment of the trainer on the trajectory.

The short duration of simulator training (20 mn) prior to cadaver testing is another questionable matter. The duration of simulator training was chosen to obtain the best operator concentration which decreased after this period. However, we believe short simulator training sessions should be performed to familiarize with this device, then anatomical theoretical knowledge could be implemented in stages via X-rays. Therefore, procedural knowledge might be acquired on a long-term basis.

The number of radiographs required by each surgeon might be considered as a debatable indicator. Such criterion was chosen since it appears as a fair reflection of the surgeon's experience in the learning curve of the iliosacral screw placement. The surgeon is more confident regarding the potential consequences of his gesture. This could therefore decrease the need for control X-rays at each stage of the wire insertion. However the analysis of this parameter gives interesting results in this study.

The iatrogenic index is questionable. No significant difference could be established between the two groups. We could have simply chosen the occurrence or not of an extra-osseous trajectory during wire positioning as standard of evaluation. We wanted to include the perception of danger by the operator according to 5 iatrogenic indicators. This parameter appeared little sensitive since the mean index did not exceed 1.8 for the whole studied population. The change from level 1 to level 2 is only due to the occurrence of an extra-osseous trajectory. Other factors should be considered such as the early occurrence of an extra-osseous trajectory during the trial session or the need for a radiographic control prior to the crossing of bone cortex. This type of advanced video analysis should be planned for better evaluation. The lack of significant difference between the iatrogenic index of N/D, T and P characteristics confirms the low sensitivity of this index. This notion appears difficult to express. The notion of risk is generally mentioned by the senior surgeon to the less experienced one in order to rectify the procedure prior to the occurrence of any complication. We do not currently have any parameter to assess it.

This study raises concerns about the assessment of a surgical simulator. We are working on the combination of a computerized virtual reality simulation with analog simulation on model. We casted a model from a synthetic skeleton inserted in a polyurethane foam. We first conducted a training session using both methods, the virtual reality simulator of the present study and the analog simulation on model. Preliminary results from 5 novice interns demonstrated a good transfer of learnings from the simulator to the model, in almost real-life conditions. Standardization of evaluation tools is easier on model than cadaver. Qualitative and quantitative analysis of radiographs appears more reproducible. This staged training from simulator to patient through the use of a model is useful when applied to the training of uncommon percutaneous surgical techniques and proves to have good time efficiency (half an hour of simulation and half an hour of practical experience with the model).

The results obtained from the number of radiographs required could be interpreted. An obvious difference has been demonstrated between G1 and G2 when novice (N) and skilled (D) operators are distinguished. N operators from G1 require the need for more radiographs than N operators from G2. The same result is found for operators with little procedural knowledge (P-). Regarding theoretical knowledge (T), T+ operators from G1 require more X-rays than T+ from G2. The use of the simulator improves the procedural knowledge of N and P-. However, for the T characteristic, this type of simulator does not seem to improve the theoretical, declarative knowledge not acquired by the surgeon. Only for the small group of operators with theoretical knowledge, a significant decrease in the number of X-rays was observed between G1 and G2. The required descriptive and topographic anatomical knowledge thus remains the fundamental basis on which the surgeon builds surgical procedures and controls.

The simulator only gives a final comment but does not intervene during the procedure. It was first designed for the familiarization with radiographs and the recognition of anatomical landmarks. According to our hypothesis, surgical simulation is a valuable training method which helps combine the useful sacroiliac joint anatomical fundaments with screw placement and specific radiographic views. We believe the decrease in the number of radiographs is indicative of a successful training. A better representation of the wire orientation in the sacrum provides higher efficiency in determining the trajectory.

N, T+ and P- are the ideal profiles for this simulator. One should possess strong anatomical theoretical knowledge and should have had little opportunity to attend an operation or to refer to a detailed operative technique.

In the orthopaedic field, the development of surgical simulators focuses on knee arthroscopy [17]. In traumatology, bone trajectory simulators are available. Some surgical implant distributors supply a simulator for the placement of cervico-cephalic screws using a handle connected to a force feedback device, such as the Melerit TraumaVision™ simulator system [www.meleritmedical.com]. A wire progresses on face and AP 2-D images of the proximal end of the femur. The radiographic views can not be chosen. This tool is presented as a marketing gadget which focuses on the device to be implanted. The feedback device gives precious information such as the crossing of the cortical bone. When wire progression is difficult but occurs too early, there is a trajectory error.

The training tool should explicitly help to mentally make the intraoperative 2-D radiograph correspond with the patient 3-D environment when applied to a percutaneous procedure. In clinical practice, errors often occur when the surgeon uses inaccurate radiographs. During iliosacral screw insertion for example, the outlet view should accurately disclose the second anterior sacral foramen on the symphysis pubis radiograph to achieve proper trajectory. Any surgical simulator should first produce 2-D images from a 3-D CT volume. The surgeon should be able to modify the radiographic views of the patient anatomy if not accurate enough.

On this basis, the simulator may offer a manual interface connected to a force feedback device to provide a better real-life situation. This tactile improvement of the interface is appealing but requires a more sophisticated equipment than a simple computerized workstation.

We particularly wish to generate data from the practical exercises performed on the simulator. Any identified and recorded error should be analysed. Blyth et al. [14] have tested a simulator for screw-plate osteosynthesis of the proximal femur in hospital students, specialized interns and trained operators. One interesting point of this simulator is the ability to obtain quantitative data from implant insertion. Qualitative data is obtained via a questionnaire. We aim to develop a simulator that would generate qualitative and quantitative data to provide a more subtle analysis of applied surgical procedures. This data will be analysed using the mathematical Bayesian approach in order to find a problem solving [18, 19]. A course, focusing on the encountered difficulty, will give the operator the knowledge he lacks to resolve the difficulty.

Another purpose of our work is to extend the design of such tools to other orthopaedic percutaneous procedures. We are working at the moment on transpedicular vertebroplasty. This procedure is more common than iliosacral screw insertion but the clinical consequences of an extra-osseous trajectory are greater, particularly in the thoracic region. Training with surgical simulator, if performed in sufficiently realistic conditions, is a valuable tool for procedural familiarization and becomes a precondition for the carrying out of these gestures in patients by a trainee surgeon.

## Conclusion

Virtual reality simulation of iliosacral screw insertion reduces the need for intraoperative radiographs during guide-wire positioning in human cadaver. Novice surgeons having a good anatomical knowledge of the lumbo-sacral joint but not used to the surgical procedure are those who will best benefit from this valuable tool. Its major educational contribution is the intraoperative use of 2-D radiographic images to guide the surgical gesture in the 3-D space. A tactile interface might be added. Performance quality can be assessed from saved data during trials which enables the surgeon to be oriented towards the corresponding lesson to be improved. These simulators are also useful in the evaluation of the operator's surgical skills prior to undertaking complex percutaneous procedures, in the spine for example.

**Figures**

Figure 1 : Abstract of the questionnaire form in order to classify surgeon skills regarding iliosacral screw placement.

The trajectory shown on inlet and outlet fluoroscopic views is wrong. Without changing the direction of the trajectory, how could we translate the entry point in order to get a correct one.

- a- To the head of the patient
- b- To the foot of the patient
- c- Ventral and to the feet of the patient, following the outlet view beam axis
- d- dorsal and to the feet of the patient, following the inlet view beam axis

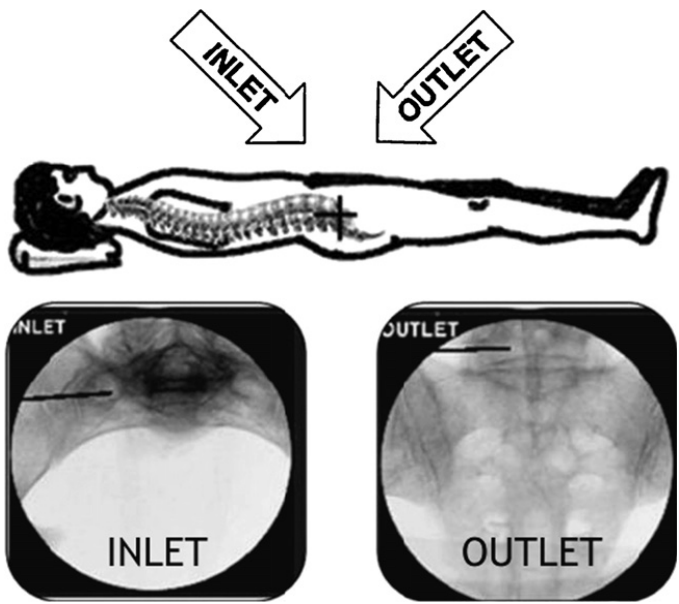

Figure 2 : simulator screen showing the CT scan 3D volume hidden by skin, X-ray views and commands

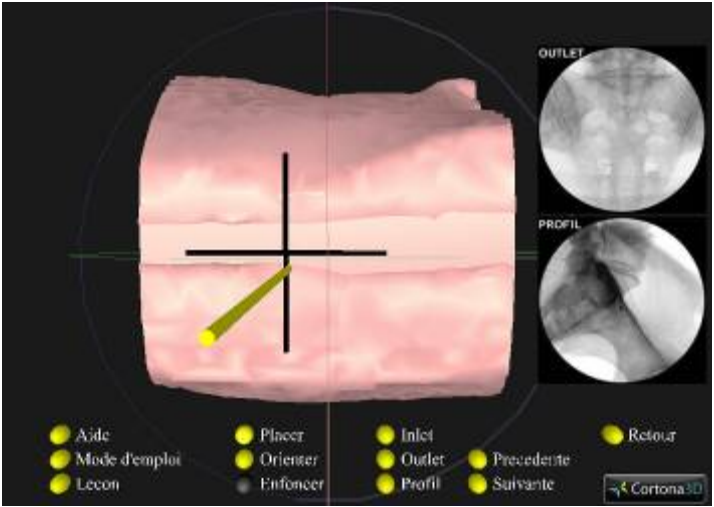